\documentclass[11pt,a4paper]{article}
\pdfoutput=1
\usepackage[a4paper, left=1.5cm, right=2cm, top=1.5cm, bottom=2.5cm]{geometry}
\usepackage[english]{babel}
\usepackage[T1]{fontenc}
\usepackage{booktabs}
\usepackage{lmodern}
\usepackage{hyperref}
\hypersetup{
    colorlinks=true,    
    linkcolor=blue,     
}

\usepackage{url,hyperref,lineno,microtype,subcaption,lscape}
\usepackage{amssymb}
\usepackage[utf8]{inputenc}
\DeclareUnicodeCharacter{0308}{\"{u}}

\usepackage{amsmath}
\usepackage{abstract}
\usepackage[onehalfspacing]{setspace}
\usepackage{csquotes}
\usepackage[style=authoryear, backend=bibtex, maxcitenames=1]{biblatex}
\DeclareCiteCommand{\textcite}
  {\boolfalse{citetracker}%
   \boolfalse{pagetracker}%
   \usebibmacro{prenote}}
  {\ifnameundef{labelname}
    {\PackageWarningNoLine{biblatex}
       {No author name available for citation}}
    {\printnames[last-first]{labelname}}}
  {\multicitedelim}
  {\usebibmacro{textcite:postnote}}

\usepackage{braket}
\usepackage{graphicx}
\usepackage{multirow}
\usepackage{float}
\usepackage{pdfpages}
\usepackage{amsfonts}
\usepackage{listings}
\usepackage{geometry}

\title{Radiative Transfer and Inversion Codes for Characterizing Planetary Atmospheres: An Overview}\date{}
\begin{document}
\maketitle

\textbf{Authors:}
    Rengel, M. \footnote{Max-Planck-Institut für Sonnensystemforschung, Justus-von-Liebig-Weg 3, 37077 Göttingen, Germany},  Adamczewski, J. $^{1, }$\footnote{Georg-August-Universität Göttingen, Friedrich-Hund-Platz 1, 37077 Göttingen, Germany} \\
    
    \textbf{Correspondence:} \texttt{rengel@mps.mpg.de}

\begin{abstract}
    The study of planetary atmospheres is crucial for understanding the origin, evolution, and processes that shape celestial bodies like planets, moons and comets. The interpretation of planetary spectra requires a detailed understanding of radiative transfer (RT) and its application through computational codes. With the advancement of observations, atmospheric modelling, and inference techniques, diverse RT and retrieval codes in planetary science have been proliferated. 
However, the selection of the most suitable code for a given problem can be challenging. To address this issue, we present a comprehensive mini-overview of the different RT and retrieval codes currently developed or available in the field of planetary atmospheres. 
This study serves as a valuable resource for the planetary science community by providing a clear and accessible list of codes, and offers a useful reference for researchers and practitioners in their selection and application of RT and retrieval codes for planetary atmospheric studies.
\end{abstract}

\section*{Introduction}
Planetary science, including the study of exoplanets, is currently a very active and fascinating multi-disciplinary field covering astronomy,  astrophysics, geophysics, astrobiology, among other fields. 
By studying planets, we can answer fundamental questions about their origin, formation, evolution, potential for life, and physico-chemical processes. The spectrum of a planet contains valuable information about its atmospheric chemical composition and physical processes. Deriving atmospheric properties like the vertical thermal structure, chemical composition and dynamics is essential to understand the origin, evolution, and how the atmospheres are influenced by physical and chemical mechanisms. 

Numerous physico-chemical problems in planetary atmospheres require a detailed understanding of radiative transfer (RT) of photons in different environments. Detailed solutions of the RT equation demand high computational capabilities. The technical implementation of these solutions in codes is a rapidly developing area, and numerous RT approaches have been attempted in recent years.  Several groups then attacked the problem from a variety of perspectives by using completely different, independent and dedicated numerical algorithms. Some atmospheric RT codes have been developed with a particular aim or for the interpretation of a particular planet. Later on, in some cases, their functionalities have been extended beyond their original purposes but still limited to the study of a specific type of planets. In recent years, breakthroughs in atmospheric retrieval have been made possible through a combination of advancements in instrumentation, computational resources, sophisticated atmospheric modelling, and powerful statistical inference techniques. 

There are varying levels of complexity in atmospheric models and the physico-chemical processes considered in RT and inversion codes, motivated by atmospheric observations. As a result, a significant number of RT and inversion codes have been developed and widely used in different atmospheric contexts, depending on whether the planet being studied is Earth, another planet or body in the Solar System, or an exoplanet. Spectra of Solar System planets can be spatially distributed and have a high signal-to-noise ratio, and can be supplemented with in situ measurements or a priori knowledge. However, exoplanetary spectroscopy currently lacks this observational capability, and retrievals must navigate a much larger parameter space for exoplanetary atmospheres than that for Solar System planets.

The selection of a particular code may depend on various factors, such as the physical problem to face, the estimation technique, computation time, flexibility, user-friendliness, among others. The selection of the most appropriate code suited for a specific problem may result in more accurate and reliable results, highlighting the importance of having a comprehensive list of available codes.
This mini-review provides a comprehensive overview of the current RT and inversion codes used in the planetary and exoplanetary communities, as seen from the perspective of a user. These codes are crucial in predicting and interpreting spectra of planetary atmospheres, both in hydrostatic equilibrium and in expanding comas. The quality and extent of these codes are critical for the effective use of space and ground-based telescopes facilities. \cite{Madhusudhan_2018} has already provided a review of exoplanetary atmospheric retrieval, some existing retrieval codes, and a description of their differences,  \cite{Barstow2020b} provides a
discussion of open problems in retrieval analysis, and more recently, \cite{MacDonald_2023} provides a catalogue of the atmospheric retrieval codes for exoplanets published to date. In this mini-review, we expand and update the codes list, including a large new generation of RT and atmospheric retrieval codes for Solar System and exoplanets.

\section*{RT and inversion codes}

One commonly used way to interpret the measured spectra is calculating a synthetic spectrum for comparison with that measured by solving the radiative transfer (forward model) -i.e., computation of the outgoing radiation from the planetary surface for a given set of free parameters-, and inferring parameters like temperature and chemical abundance profiles. This last step is called inversion or retrieval and consists in 
comparing the measured and best modelled/synthetic spectra adjusting the atmospheric parameters in such a way as to minimise any discrepancy. A number of radiative transfer codes or forward models and inversion algorithms are already constructed, and generally available and used by the planetary and exoplanetary characterisation communities. A comprehensive list of RT and retrieval codes in the literature is shown
in Table \ref{table:1}, indicating where to find them (if available) and the estimation technique employed (Sect. 3).

Retrieval codes can use either parametric models or self-consistent equilibrium models to estimate the composition and pressure-temperature (P-T) profiles from spectral data. Parametric models do not make any assumptions about thermochemical and radiative-convective equilibrium, using instead parametric forms for the P-T profile and composition. Self-consistent models compute profiles based on assumptions of the atmosphere's physical and chemical properties and processes, with varying levels of complexity from 1-D atmospheres to full 3-D general circulation models. We do not cover 3-D general circulation codes specifically in this paper, they are complex and couple many processes together in a three-dimensional time-marching calculation.
For a discussion of the principles of atmospheric retrievals of exoplanets see \cite{Gandhi_2017}. When calculating a synthetic atmospheric spectrum, lines are read from atomic and molecular databases -databases differ in completeness and accuracies-, and opacities are calculated via diverse approaches 
\cite{rengel_2022b}. 
 
 Some RT models used for Solar System planets and exoplanets have their roots in Earth codes. There are several RT codes used in the Earth atmospheric community but they are just listed here: 4A/OP [\cite{Scott_1974, Scott_1981}], 6S [\cite{Lee_2019}], AccuRT\footnote{\url{http://www.geminor.com/accurt.html}} [\cite{Stamnes_2018}], ARMS \cite{Weng_2020}, BIRRA\footnote{\url{https://www.dlr.de/eoc/en/desktopdefault.aspx/tabid-5447/9970_read-20673/}} [\cite{Gimeno_2011}], BTRAM\footnote{\url{https://blueskyspectroscopy.com/?page_id=21}} [\cite{Chapman_2004}], DISORT [\cite{Stamnes_1988}], Eradiate\footnote{\url{https://www.eradiate.eu/site/}} [\cite{Govaerts_2022}], FASCODE [\cite{Smith_1978, Clough_1981}], FUTBOLIN [\cite{Martin-Torres_2005}], GARLIC [\cite{Schreier_2014}], GENLN2 [\cite{Edwards_1992}], IMAP-DOAS [\cite{Frankenberg_2005}], IMLM \cite{Gloudemans_2005}, KCARTA\footnote{\url{http://asl.umbc.edu/pub/packages/kcarta.html}},\footnote{\url{https://github.com/sergio66/kcarta\_gen}} [\cite{DeSouza-Machado_2020}], LBLRTM\footnote{\url{http://rtweb.aer.com/lblrtm.html}},\footnote{\url{https://github.com/AER-RC/LBLRTM}} [\cite{Clough_2005}], LEEDR [\cite{Fiorino_2014}], LinePak\footnote{\url{https://spectralcalc.com/info/about}} [\cite{Gordley_1994}], libRadtran\footnote{\url{http://www.libradtran.org/doku.php}} [\cite{Mayer_2005}], MODTRAN\footnote{\url{http://modtran.spectral.com}} [\cite{Berk_1998}], MPI\footnote{\url{https://mitpress.mit.edu/9780262571234/mpi-the-complete-reference/}} [\cite{Gropp_1998}], PUMAS\footnote{\url{https://psg.gsfc.nasa.gov/}} [\cite{Villanueva_2015, Villanueva_2022}], RFM\footnote{\url{http://eodg.atm.ox.ac.uk/RFM/}} [\cite{Dudhia_2017}], RTMOM [\cite{Govaerts_2006}], RTTOV\footnote{\url{https://nwp-saf.eumetsat.int/site/software/rttov/}} [\cite{Saunders_1999, Matricardi_2004}], SARTA\footnote{\url{https://github.com/strow/sarta}}\footnote{\url{https://github.com/clhepp/sarta}} [\cite{Strow_2003}], SARTre [\cite{mendrok2006sartre}], SASKTRAN\footnote{\url{https://arg.usask.ca/docs/sasktran/}} [\cite{Bourassa_2008, Zawada_2015}], SBDART\footnote{\url{https://github.com/paulricchiazzi/SBDART}} [\cite{Ricchiazzi_2002}], SCIATRAN\footnote{\url{https://www.iup.uni-bremen.de/sciatran/}} [\cite{Rozanov_2005, Rozanov_2014}], SICOR\footnote{\url{https://pypi.org/project/sicor/}} [\cite{Borsdorff_2017, Borsdorff_2018}], SKIRT\footnote{\url{https://skirt.ugent.be/root/\_home.html}} [\cite{Baes_2003}], SMART-G\footnote{\url{https://www.hygeos.com/smartg}} [\cite{Ramon_2019}], TOMRAD [\cite{Dave_1964}], VDISORT [\cite{Lin_2022}], VLIDORT/LIDORT\footnote{\url{http://www.rtslidort.com/about\_overview.html}} [\cite{Spurr_2019}], WFM-DOAS [\cite{Buchwitz_2004, Buchwitz_2005}] as well as models by \cite{Hartogh_1989}, \cite{Tinetti_2006a, Tinetti_2006b} and \cite{Robinson_2011}.


\section*{Retrieval or parameter fitting or estimation techniques}

\begin{figure}[h!]
\begin{center}
\includegraphics[width=16cm]{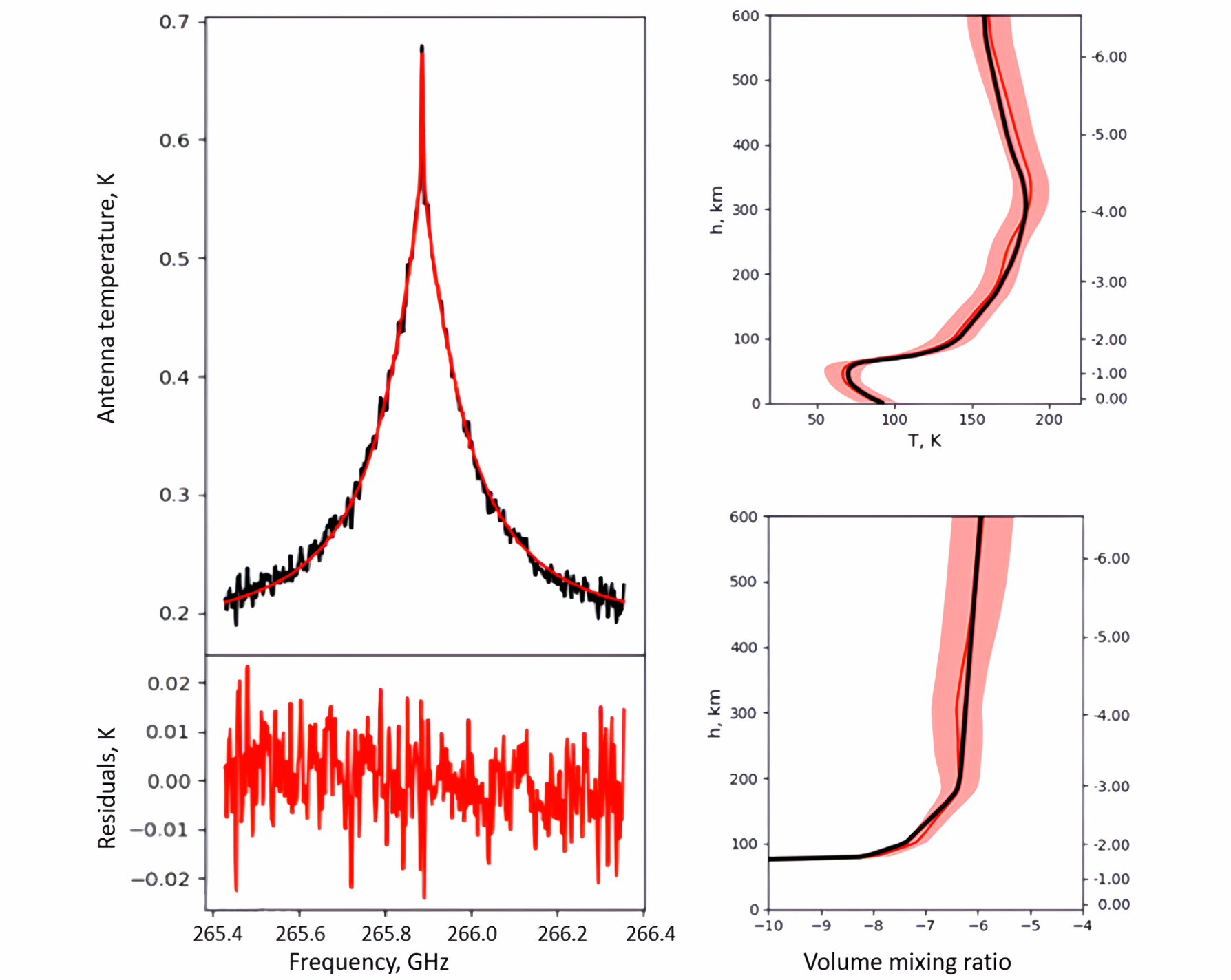}
\end{center}
\caption{Example of forward model and retrieval applied to the case of HCN in Titan: Left: comparison between observed and best-fit simulated HCN (3-2) lines (black and red, respectively, upper panel), and the difference between the observed and fitted spectra (lower panel). Right: retrieved temperature and HCN distribution derived from the spectrum. The black and red lines, and the pink shadow show the initial and retrieved profiles, and the error bars, respectively. Figure based on Fig. 1 in \cite{Rengel_2022}}.\label{fig:1}
\end{figure}

The objective of an optimization algorithm is to extensively and efficiently sample a high-dimensional parameter space to find the best solution space given the data. The retrieval or parameter fitting or estimation techniques commonly used are Optimal Estimation (OE) algorithm ($\dagger$), nested sampling ($\ddagger$), Markov chain Monte Carlo (MCMC) method ($\blacktriangle$), Grid search ($\bullet$), and more recently, data-driven Machine Learning (ML) approaches ($\blacklozenge$). OE is popular in the Solar System community and assumes Gaussian statistics. It is fast and efficient and applicable to exoplanets under some specific conditions (e.g. \cite{Rengel_2008,Hartogh_2010,Lee_2011,Line_2012,Shulyak_2019,Rengel_2022,Villanueva_2022}. Grid search is simple and computationally cheap, but it can be inefficient (e.g. Madhusudhan \& Seager 2009). MCMC provides a better parameter exploration of the parameter space but with limitations in calculating the model evidence for model comparison and can be computationally expensive (e.g. \cite{Benneke_2012,Madhusudhan_2014,Waldmann_2015a,Blecic_2016,Cubillos_2016,Evans_2017,Wakeford_2017,Lacy_2020}). Nested sampling algorithm facilitates efficient parameter space exploration and calculation of model evidence (e.g. \cite{Benneke_2013,Waldmann_2015b,Gandhi_2017,Pinhas_2018,Mollière_2019,Brogi_2019,Fisher_2019,Shulyak_2020,Min_2020,Seidel_2020}). ML algorithms can be computationally efficient, but they require large amounts of training data and can be sensitive to biases in the training set (e.g. \cite{Waldmann_2016,Marquez-Neila_2018,Zingales_2018,Soboczenski_2018,Cobb_2019,Fisher_2020,Nixon_2020,Hayes_2020}). An example of an application of the forward model and retrieval technique is provided in Figure \ref{fig:1}.


\section*{Verification and validation of RT and retrieval codes}

The retrieval problem is a challenging one due to its ill-posed and ill-conditioned nature. This means that even a small change in measurements can result in a significant deviation in the estimated model, making the inverse solution computation extremely unstable (e.g. \cite{Ih_2021}). Currently, each retrieval code has its unique method for computing opacities and input models, lacking community standards. Furthermore, models continue to become more complex and data quality improves, 
retrieval codes face increasingly complex and degenerate problems (e.g. \cite{Welbanks_2019}).

Verification and validation of planetary retrieval codes is an important aspect. Verification can be readily accomplished using synthetic measurements and code inter-comparison. Inter-model comparisons of forward and retrieval suites have already been underway, with notable studies by \cite{Clarmann_2003b,Baudino_2017, Schreier_2018, Barstow_2020, Barstow_2022}.
Validation is challenging due to the lack of reference “truth” data (e.g. in situ measurements). Thus,  for an assessment of exoplanet atmospheric remote sensing, data dedicated to solar system planets like Earth, Mars and Venus is an ideal test case. To our knowledge, this kind of data have been rarely used to demonstrate the capabilities of exoplanet atmospheric studies. Observing planetary transits in our own Solar System for example can serve as an invaluable benchmark and can provide crucial information for future exoplanet characterizations (e.g. \cite{Ehrenreich_2011, Ehrenreich_2012, Montanes-Rodriguez_2015, Lopez-Puerats_2018, Schreier_2018}).\\


Results from inter-comparison show that that small differences in the forward model setup can lead to noticeable differences in the retrieval outcome (\cite{Barstow_2020}). These efforts are crucial for improving the field and providing a clearer understanding of the sources of variability and their impact on the results. Such comparisons provide a roadmap for future advancements and refinements in modelling, enabling us to compare theoretical predictions with observations. They also offer an opportunity to identify any remaining problems, leading to the improvement of existing codes and the development of more reliable and consistent codes. The continued pursuit of inter-model comparisons will drive the field forward towards a better understanding of the retrieval problem. Furthermore, the inherent similarities in methods was investigated (\cite{Line_2013}), and there is now
considerable work to understand and quantify the role of different modelling choices affecting accuracy
and precision on constraints (\cite{Barstow_2020}).

\section*{Final remarks}
Our current understanding of planetary atmospheres is
limited by model completeness and robustness. Forward modelling and spectral retrieval is the leading technique for interpretation of spectra and is employed by various teams using a variety of forward models and parameter estimation algorithms.
Numerous codes are available and continue been developed. The code employed which contain the complexity of models should be chosen with care depending on the specific questions whose answer is sought and on the nature of the data at hand (considering spectral coverage and region, resolution, precision, signal-to-noise ratio, etc.). The efficiency of these codes is a critical issue, but beyond the scope of this paper. 

The study of RT in planetary atmospheres is a rapidly developing field, with numerous advances being made in both the understanding of the RT process and in the development of RT codes. With continued research and development including the improving of observational capabilities, it is likely that even more sophisticated RT codes and inversion algorithms will be developed in the future, enabling us to better understand the diversity of planetary systems beyond our own and the mechanisms shaping their atmospheres.


\section*{Author Contributions}
MR wrote the first draft of the manuscript. All authors contributed to manuscript revision, read, and approved the submitted version.

\section*{Funding}
Authors acknowledge the support by the DFG priority program SPP 1992 “Exploring the Diversity of Extrasolar Planets” (DFG PR 36 24602/41). 

\section*{Acknowledgments}
This research has made use of NASA's Astrophysics Data System Bibliographic Services.

\printbibliography[heading=bibintoc, title={References}]

@phdthesis{mendrok2006sartre,
  title={The SARTre model for radiative transfer in spherical atmospheres and its application to the derivation of cirrus cloud properties},
  author={Mendrok, J.},
  year={2006},
  school={Freie Univ., Berlin}
}

@techreport{stamnes2000,
  author = {Stamnes, K. and Tsay, S. C. and Wiscombe, W. J. and Laszlo, I.},
  title = {{DISORT: A General-Purpose Fortran Program for Discrete Ordinate-Method Radiative Transfer in Scattering and Emitting Layered Media: Documentation of Methodology}},
  year = {2000},
  institution = {NASA},
  address = {Greenbelt, United States},
  type = {NASA Report},
  url = {ftp://climate1.gsfc.nasa.gov/wiscombe/Multiple_Scatt/DISORTReport1.1.pdf}
}

@ARTICLE{Barstow2020b,
       author = {{Barstow}, Joanna K. and {Heng}, Kevin},
        title = "{Outstanding Challenges of Exoplanet Atmospheric Retrievals}",
      journal = {ssr},
         year = 2020,
        month = jun,
       volume = {216},
       number = {5},
          eid = {82},
        pages = {82},
          doi = {10.1007/s11214-020-00666-x},
archivePrefix = {arXiv},
       eprint = {2003.14311},
 primaryClass = {astro-ph.EP},
       adsurl = {https://ui.adsabs.harvard.edu/abs/2020SSRv..216...82B}
}

@Article{Stamnes_2018,
AUTHOR = {Stamnes, Knut and Hamre, Børge and Stamnes, Snorre and Chen, Nan and Fan, Yongzhen and Li, Wei and Lin, Zhenyi and Stamnes, Jakob},
TITLE = {Progress in Forward-Inverse Modeling Based on Radiative Transfer Tools for Coupled Atmosphere-Snow/Ice-Ocean Systems: A Review and Description of the AccuRT Model},
JOURNAL = {Applied Sciences},
VOLUME = {8},
YEAR = {2018},
NUMBER = {12},
ARTICLE-NUMBER = {2682},
URL = {https://www.mdpi.com/2076-3417/8/12/2682},
ISSN = {2076-3417},
DOI = {10.3390/app8122682}
}

@article{MacDonald_2023,
	doi = {10.3847/2515-5172/acc46a},
    
	year = 2023,
	month = {mar},
  
	publisher = {American Astronomical Society},
  
	volume = {7},
  
	number = {3},
  
	pages = {54},
  
	author = {Ryan J. MacDonald and Natasha E. Batalha},
  
	title = {A Catalog of Exoplanet Atmospheric Retrieval Codes},
  
	journal = {Research Notes of the {AAS}
}
}

@article{Lupu_2016,
doi = {10.3847/0004-6256/152/6/217},
url = {https://dx.doi.org/10.3847/0004-6256/152/6/217},
year = {2016},
month = {dec},
publisher = {The American Astronomical Society},
volume = {152},
number = {6},
pages = {217},
author = {Roxana E. Lupu and Mark S. Marley and Nikole Lewis and Michael Line and Wesley A. Traub and Kevin Zahnle},
title = {DEVELOPING ATMOSPHERIC RETRIEVAL METHODS FOR DIRECT IMAGING SPECTROSCOPY OF GAS GIANTS IN REFLECTED LIGHT. I. METHANE ABUNDANCES AND BASIC CLOUD PROPERTIES},
journal = {The Astronomical Journal},

}

@article{Howe_2022,
doi = {10.3847/1538-4357/ac5590},
url = {https://dx.doi.org/10.3847/1538-4357/ac5590},
year = {2022},
month = {aug},
publisher = {The American Astronomical Society},
volume = {935},
number = {2},
pages = {107},
author = {Alex R. Howe and Michael W. McElwain and Avi M. Mandell},
title = {GJ 229B: Solving the Puzzle of the First Known T Dwarf with the APOLLO Retrieval Code},
journal = {The Astrophysical Journal},
}

@article{Howe_2017,
doi = {10.3847/1538-4357/835/1/96},
url = {https://dx.doi.org/10.3847/1538-4357/835/1/96},
year = {2017},
month = {jan},
publisher = {The American Astronomical Society},
volume = {835},
number = {1},
pages = {96},
author = {Alex R. Howe and Adam Burrows and Drake Deming},
title = {AN INFORMATION-THEORETIC APPROACH TO OPTIMIZE JWST OBSERVATIONS AND RETRIEVALS OF TRANSITING EXOPLANET ATMOSPHERES},
journal = {The Astrophysical Journal},
}

@article{Burningham_2017,
    author = {Burningham, Ben and Marley, M. S. and Line, M. R. and Lupu, R. and Visscher, C. and Morley, C. V. and Saumon, D. and Freedman, R.},
    title = "{Retrieval of atmospheric properties of cloudy L dwarfs}",
    journal = {Monthly Notices of the Royal Astronomical Society},
    volume = {470},
    number = {1},
    pages = {1177-1197},
    year = {2017},
    month = {05},
    issn = {0035-8711},
    doi = {10.1093/mnras/stx1246},
    url = {https://doi.org/10.1093/mnras/stx1246},
    eprint = {https://academic.oup.com/mnras/article-pdf/470/1/1177/17933870/stx1246.pdf},
}

@article{espinoza_2018,
    author = {Espinoza, Néstor and Rackham, Benjamin V and Jordán, Andrés and Apai, Dániel and López-Morales, Mercedes and Osip, David J and Grimm, Simon L and Hoeijmakers, Jens and Wilson, Paul A and Bixel, Alex and McGruder, Chima and Rodler, Florian and Weaver, Ian and Lewis, Nikole K and Fortney, Jonathan J and Fraine, Jonathan},
    title = "{ACCESS: a featureless optical transmission spectrum for WASP-19b from Magellan/IMACS}",
    journal = {Monthly Notices of the Royal Astronomical Society},
    volume = {482},
    number = {2},
    pages = {2065-2087},
    year = {2018},
    month = {10},
    issn = {0035-8711},
    doi = {10.1093/mnras/sty2691},
    url = {https://doi.org/10.1093/mnras/sty2691},
    eprint = {https://academic.oup.com/mnras/article-pdf/482/2/2065/26445094/sty2691.pdf},
}

@ARTICLE{Weng_2020,
       author = {{Weng}, Fuzhong and {Yu}, Xinwen and {Duan}, Yihong and {Yang}, Jun and {Wang}, Jianjie},
        title = "{Advanced Radiative Transfer Modeling System (ARMS): A New-Generation Satellite Observation Operator Developed for Numerical Weather Prediction and Remote Sensing Applications}",
      journal = {Advances in Atmospheric Sciences},
         year = 2020,
        month = jan,
       volume = {37},
       number = {2},
        pages = {131-136},
          doi = {10.1007/s00376-019-9170-2},
}

@article{Stolker_2020,
	author = {{Stolker}, T. and {Quanz, S. P.} and {Todorov, K. O.} and {K\"uhn, J.} and {Molli\`ere, P.} and {Meyer, M. R.} and {Currie, T.} and {Daemgen, S.} and {Lavie, B.}},
	title = {MIRACLES: atmospheric characterization of directly imaged planets and substellar companions at 4-5  - I. Photometric analysis of  b, HIP 65426 b, PZ Tel B, and HD 206893 B},
	DOI= "10.1051/0004-6361/201937159",
	url= "https://doi.org/10.1051/0004-6361/201937159",
	journal = {A\&A},
	year = 2020,
	volume = 635,
	pages = "A182",
}

@article{Gibson_2020,
    author = {Gibson, Neale P and Merritt, Stephanie and Nugroho, Stevanus K and Cubillos, Patricio E and de Mooij, Ernst J W and Mikal-Evans, Thomas and Fossati, Luca and Lothringer, Joshua and Nikolov, Nikolay and Sing, David K and Spake, Jessica J and Watson, Chris A and Wilson, Jamie},
    title = "{Detection of Fe I in the atmosphere of the ultra-hot Jupiter WASP-121b, and a new likelihood-based approach for Doppler-resolved spectroscopy}",
    journal = {Monthly Notices of the Royal Astronomical Society},
    volume = {493},
    number = {2},
    pages = {2215-2228},
    year = {2020},
    month = {01},
    issn = {0035-8711},
    doi = {10.1093/mnras/staa228},
    url = {https://doi.org/10.1093/mnras/staa228},
    eprint = {https://academic.oup.com/mnras/article-pdf/493/2/2215/32755559/staa228.pdf},
}

@article{Damiano_2020,
doi = {10.3847/1538-3881/ab79a5},
url = {https://dx.doi.org/10.3847/1538-3881/ab79a5},
year = {2020},
month = {mar},
publisher = {The American Astronomical Society},
volume = {159},
number = {4},
pages = {175},
author = {Mario Damiano and Renyu Hu},
title = {ExoReL: A Bayesian Inverse Retrieval Framework for Exoplanetary Reflected Light Spectra},
journal = {The Astronomical Journal},
}

@article{Changeat_2020,
doi = {10.3847/1538-3881/ab9a53},
url = {https://dx.doi.org/10.3847/1538-3881/ab9a53},
year = {2020},
month = {jul},
publisher = {The American Astronomical Society},
volume = {160},
number = {2},
pages = {80},
author = {Q. Changeat and A. Al-Refaie and L. V. Mugnai and B. Edwards and I. P. Waldmann and E. Pascale and G. Tinetti},
title = {Alfnoor: A Retrieval Simulation of the Ariel Target List},
journal = {The Astronomical Journal},
}

@article{Carrion-Gonzalez_2020,
	author = {{Carri\'on-Gonz\'alez}, \'O. and {Garc\'{\i}a Mu\~noz, A.} and {Cabrera, J.} and {Csizmadia, Sz.} and {Santos, N. C.} and {Rauer, H.}},
	title = {Directly imaged exoplanets in reflected starlight: the importance of knowing the planet radius},
	DOI= "10.1051/0004-6361/202038101",
	url= "https://doi.org/10.1051/0004-6361/202038101",
	journal = {A\&A},
	year = 2020,
	volume = 640,
	pages = "A136",
}

@article{Swain_2014,
doi = {10.1088/0004-637X/784/2/133},
url = {https://dx.doi.org/10.1088/0004-637X/784/2/133},
year = {2014},
month = {mar},
publisher = {The American Astronomical Society},
volume = {784},
number = {2},
pages = {133},
author = {Mark R. Swain and Michael R. Line and Pieter Deroo},
title = {ON THE DETECTION OF MOLECULES IN THE ATMOSPHERE OF HD 189733b USING HST NICMOS TRANSMISSION SPECTROSCOPY},
journal = {The Astrophysical Journal},
}

@article{Kawahara_2022,
doi = {10.3847/1538-4365/ac3b4d},
url = {https://dx.doi.org/10.3847/1538-4365/ac3b4d},
year = {2022},
month = {jan},
publisher = {The American Astronomical Society},
volume = {258},
number = {2},
pages = {31},
author = {Hajime Kawahara and Yui Kawashima and Kento Masuda and Ian J. M. Crossfield and Erwan Pannier and Dirk van den Bekerom},
title = {Autodifferentiable Spectrum Model for High-dispersion Characterization of Exoplanets and Brown Dwarfs},
journal = {The Astrophysical Journal Supplement Series},
}

@article{Challener_2022,
doi = {10.3847/1538-3881/ac4885},
url = {https://dx.doi.org/10.3847/1538-3881/ac4885},
year = {2022},
month = {feb},
publisher = {The American Astronomical Society},
volume = {163},
number = {3},
pages = {117},
author = {Ryan C. Challener and Emily Rauscher},
title = {ThERESA: Three-dimensional Eclipse Mapping with Application to Synthetic JWST Data},
journal = {The Astronomical Journal},
}

@article{Dos_Santos_2022,
	author = {{Dos Santos}, Leonardo A. and {Vidotto, Aline A.} and {Vissapragada, Shreyas} and {Alam, Munazza K.} and {Allart, Romain} and {Bourrier, Vincent} and {Kirk, James} and {Seidel, Julia V.} and {Ehrenreich, David}},
	title = {
p-winds: An open-source Python code to model planetary outflows and upper atmospheres},
	DOI= "10.1051/0004-6361/202142038",
	url= "https://doi.org/10.1051/0004-6361/202142038",
	journal = {A\&A},
	year = 2022,
	volume = 659,
	pages = "A62",
}

@article{Lustig-Yaeger_2022,
doi = {10.3847/1538-3881/ac5034},
url = {https://dx.doi.org/10.3847/1538-3881/ac5034},
year = {2022},
month = {feb},
publisher = {The American Astronomical Society},
volume = {163},
number = {3},
pages = {140},
author = {Jacob Lustig-Yaeger and Kristin S. Sotzen and Kevin B. Stevenson and Rodrigo Luger and Erin M. May and L. C. Mayorga and Kathleen Mandt and Noam R. Izenberg},
title = {Hierarchical Bayesian Atmospheric Retrieval Modeling for Population Studies of Exoplanet Atmospheres: A Case Study on the Habitable Zone},
journal = {The Astronomical Journal},
}

@article{Robinson_2023,
doi = {10.3847/PSJ/acac9a},
url = {https://dx.doi.org/10.3847/PSJ/acac9a},
year = {2023},
month = {jan},
publisher = {The American Astronomical Society},
volume = {4},
number = {1},
pages = {10},
author = {Tyler D. Robinson and Arnaud Salvador},
title = {Exploring and Validating Exoplanet Atmospheric Retrievals with Solar System Analog Observations},
journal = {The Planetary Science Journal},
}

@ARTICLE{Lin_2022,
  
AUTHOR={Lin, Zhenyi and Stamnes, Snorre and Li, Wei and Hu, Yongxiang and Laszlo, Istvan and Tsay, Si-Chee and Berk, Alexander and van den Bosch, Jeannette and Stamnes, Knut},   
	 
TITLE={Polarized Radiative Transfer Simulations: A Tutorial Review and Upgrades of the Vector Discrete Ordinate Radiative Transfer Computational Tool},      
	
JOURNAL={Frontiers in Remote Sensing},      
	
VOLUME={3},           
	
YEAR={2022},      
	  
URL={https://www.frontiersin.org/articles/10.3389/frsen.2022.880768},       
	
DOI={10.3389/frsen.2022.880768},      
	
ISSN={2673-6187},   
   
}

@ARTICLE{Madhusudhan2010,
       author = {{Madhusudhan}, N. and {Seager}, S.},
        title = "{On the Inference of Thermal Inversions in Hot Jupiter Atmospheres}",
      journal = {apj},
         year = 2010,
        month = dec,
       volume = {725},
       number = {1},
        pages = {261-274},
          doi = {10.1088/0004-637X/725/1/261},
archivePrefix = {arXiv},
       eprint = {1010.4585},
 primaryClass = {astro-ph.EP},
       adsurl = {https://ui.adsabs.harvard.edu/abs/2010ApJ...725..261M}
}

@ARTICLE{Madhusudhan_2011,
       author = {{Madhusudhan}, Nikku and {Harrington}, Joseph and {Stevenson}, Kevin B. and {Nymeyer}, Sarah and {Campo}, Christopher J. and {Wheatley}, Peter J. and {Deming}, Drake and {Blecic}, Jasmina and {Hardy}, Ryan A. and {Lust}, Nate B. and {Anderson}, David R. and {Collier-Cameron}, Andrew and {Britt}, Christopher B.~T. and {Bowman}, William C. and {Hebb}, Leslie and {Hellier}, Coel and {Maxted}, Pierre F.~L. and {Pollacco}, Don and {West}, Richard G.},
        title = "{A high C/O ratio and weak thermal inversion in the atmosphere of exoplanet WASP-12b}",
      journal = {nat},
         year = 2011,
        month = jan,
       volume = {469},
       number = {7328},
        pages = {64-67},
          doi = {10.1038/nature09602},
archivePrefix = {arXiv},
       eprint = {1012.1603},
 primaryClass = {astro-ph.EP},
       adsurl = {https://ui.adsabs.harvard.edu/abs/2011Natur.469...64M}
}

@ARTICLE{Madhusudhan_2014,
       author = {{Madhusudhan}, Nikku and {Crouzet}, Nicolas and {McCullough}, Peter R. and {Deming}, Drake and {Hedges}, Christina},
        title = "{H$_{2}$O Abundances in the Atmospheres of Three Hot Jupiters}",
      journal = {apjl},
         year = 2014,
        month = aug,
       volume = {791},
       number = {1},
          eid = {L9},
        pages = {L9},
          doi = {10.1088/2041-8205/791/1/L9},
archivePrefix = {arXiv},
       eprint = {1407.6054},
 primaryClass = {astro-ph.EP},
       adsurl = {https://ui.adsabs.harvard.edu/abs/2014ApJ...791L...9M}
}

@ARTICLE{Hayes_2020,
       author = {{Hayes}, J.~J.~C. and {Kerins}, E. and {Awiphan}, S. and {McDonald}, I. and {Morgan}, J.~S. and {Chuanraksasat}, P. and {Komonjinda}, S. and {Sanguansak}, N. and {Kittara}, P. and {SPEARNet Collaboration}},
        title = "{Optimizing exoplanet atmosphere retrieval using unsupervised machine-learning classification}",
      journal = {mnras},
         year = 2020,
        month = may,
       volume = {494},
       number = {3},
        pages = {4492-4508},
          doi = {10.1093/mnras/staa978},
archivePrefix = {arXiv},
       eprint = {1909.00718},
 primaryClass = {astro-ph.EP},
       adsurl = {https://ui.adsabs.harvard.edu/abs/2020MNRAS.494.4492H}
}

@ARTICLE{Waldmann_2016,
       author = {{Waldmann}, I.~P.},
        title = "{Dreaming of Atmospheres}",
      journal = {apj},
         year = 2016,
        month = apr,
       volume = {820},
       number = {2},
          eid = {107},
        pages = {107},
          doi = {10.3847/0004-637X/820/2/107},
archivePrefix = {arXiv},
       eprint = {1511.08339},
 primaryClass = {astro-ph.EP},
       adsurl = {https://ui.adsabs.harvard.edu/abs/2016ApJ...820..107W}
}

@ARTICLE{Nixon_2020,
       author = {{Nixon}, Matthew C. and {Madhusudhan}, Nikku},
        title = "{Assessment of supervised machine learning for atmospheric retrieval of exoplanets}",
      journal = {mnras},
         year = 2020,
        month = jul,
       volume = {496},
       number = {1},
        pages = {269-281},
          doi = {10.1093/mnras/staa1150},
archivePrefix = {arXiv},
       eprint = {2004.10755},
 primaryClass = {astro-ph.EP},
       adsurl = {https://ui.adsabs.harvard.edu/abs/2020MNRAS.496..269N}
}

@ARTICLE{Scheucher2020,
       author = {{Scheucher}, Markus and {Wunderlich}, F. and {Grenfell}, J.~L. and {Godolt}, M. and {Schreier}, F. and {Kappel}, D. and {Haus}, R. and {Herbst}, K. and {Rauer}, H.},
        title = "{Consistently Simulating a Wide Range of Atmospheric Scenarios for K2-18b with a Flexible Radiative Transfer Module}",
      journal = {apj},
         year = 2020,
        month = jul,
       volume = {898},
       number = {1},
          eid = {44},
        pages = {44},
          doi = {10.3847/1538-4357/ab9084},
archivePrefix = {arXiv},
       eprint = {2005.02114},
 primaryClass = {astro-ph.EP},
       adsurl = {https://ui.adsabs.harvard.edu/abs/2020ApJ...898...44S}
}

@Article{article,
	author = "Author1 LastName1 and Author2 LastName2 and Author3 LastName3",
	title = "Article Title",
	volume = "30",
	number = "30",
	pages = "10127-10134",
	year = "2013",
	doi = "10.3389/fnins.2013.12345",
	URL = "http://www.frontiersin.org/Journal/10.3389/fnins.2013.12345/abstract",
	journal = "Frontiers in Neuroscience"
}

@Inbook{Spurr_2019,
author="Spurr, Robert
and Christi, Matt",
editor="Kokhanovsky, Alexander",
title="The LIDORT and VLIDORT Linearized Scalar and Vector Discrete Ordinate Radiative Transfer Models: Updates in the Last 10 Years",
bookTitle="Springer Series in Light Scattering: Volume 3: Radiative Transfer and Light Scattering",
year="2019",
publisher="Springer International Publishing",
address="Cham",
pages="1--62",
isbn="978-3-030-03445-0",
doi="10.1007/978-3-030-03445-0_1",
}

@article{Rozanov_2014,
title = {Radiative transfer through terrestrial atmosphere and ocean: Software package SCIATRAN},
journal = {Journal of Quantitative Spectroscopy and Radiative Transfer},
volume = {133},
pages = {13-71},
year = {2014},
issn = {0022-4073},
doi = {https://doi.org/10.1016/j.jqsrt.2013.07.004},
url = {https://www.sciencedirect.com/science/article/pii/S0022407313002872},
author = {V.V. Rozanov and A.V. Rozanov and A.A. Kokhanovsky and J.P. Burrows},
}

@article{Ramon_2019,
title = {Modeling polarized radiative transfer in the ocean-atmosphere system with the GPU-accelerated SMART-G Monte Carlo code},
journal = {Journal of Quantitative Spectroscopy and Radiative Transfer},
volume = {222-223},
pages = {89-107},
year = {2019},
issn = {0022-4073},
doi = {https://doi.org/10.1016/j.jqsrt.2018.10.017},
url = {https://www.sciencedirect.com/science/article/pii/S002240731830400X},
author = {Didier Ramon and François Steinmetz and Dominique Jolivet and Mathieu Compiègne and Robert Frouin},
}

@article{Rozanov_2005,
title = {SCIATRAN 2.0 – A new radiative transfer model for geophysical applications in the 175–2400nm spectral region},
journal = {Advances in Space Research},
volume = {36},
number = {5},
pages = {1015-1019},
year = {2005},
issn = {0273-1177},
doi = {https://doi.org/10.1016/j.asr.2005.03.012},
url = {https://www.sciencedirect.com/science/article/pii/S0273117705002887},
author = {A. Rozanov and V. Rozanov and M. Buchwitz and A. Kokhanovsky and J.P. Burrows},
}

@ARTICLE{Lopez-Puerats_2018,
       author = {{L{\'o}pez-Puertas}, Manuel and {Monta{\~n}{\'e}s-Rodr{\'\i}guez}, Pilar and {Pall{\'e}}, E. and {H{\"o}pfner}, M. and {S{\'a}nchez-L{\'o}pez}, A. and {Garc{\'\i}a-Comas}, M. and {Funke}, B.},
        title = "{Aerosols and Water Ice in Jupiter{\textquoteright}s Stratosphere from UV-NIR Ground-based Observations}",
      journal = {aj},
         year = 2018,
        month = oct,
       volume = {156},
       number = {4},
          eid = {169},
        pages = {169},
          doi = {10.3847/1538-3881/aadcef},
}

@ARTICLE{Montanes-Rodriguez_2015,
       author = {{Monta{\~n}{\'e}s-Rodr{\'\i}guez}, Pilar and {Gonz{\'a}lez-Merino}, B. and {Pall{\'e}}, E. and {L{\'o}pez-Puertas}, Manuel and {Garc{\'\i}a-Melendo}, E.},
        title = "{Jupiter as an Exoplanet: UV to NIR Transmission Spectrum Reveals Hazes, a Na Layer, and Possibly Stratospheric H$_{2}$O-ice Clouds}",
      journal = {apjl},
         year = 2015,
        month = mar,
       volume = {801},
       number = {1},
          eid = {L8},
        pages = {L8},
          doi = {10.1088/2041-8205/801/1/L8},
archivePrefix = {arXiv},
       eprint = {1502.02132},
}

@ARTICLE{Ehrenreich_2012,
       author = {{Ehrenreich}, D. and {Vidal-Madjar}, A. and {Widemann}, T. and {Gronoff}, G. and {Tanga}, P. and {Barth{\'e}lemy}, M. and {Lilensten}, J. and {Lecavelier Des Etangs}, A. and {Arnold}, L.},
        title = "{Transmission spectrum of Venus as a transiting exoplanet}",
      journal = {aap},
         year = 2012,
        month = jan,
       volume = {537},
          eid = {L2},
        pages = {L2},
          doi = {10.1051/0004-6361/201118400},
archivePrefix = {arXiv},
       eprint = {1112.0572},
}

@ARTICLE{Ehrenreich_2011,
       author = {{Ehrenreich}, D. and {Vidal-Madjar}, A. and {Widemann}, T. and {Grono}, G. and {Tanga}, P. and {Barthelemy}, M. and {Lilensten}, J. and {Lecavelier Des Etangs}, A. and {Arnold}, L.},
        title = "{VizieR Online Data Catalog: Transmission spectrum of Venus (Ehrenreich+, 2012)}",
      journal = {VizieR Online Data Catalog},
         year = 2011,
        month = nov,
          eid = {J/A+A/537/L2},
        pages = {J/A+A/537/L2},
       adsurl = {https://ui.adsabs.harvard.edu/abs/2011yCat..35379002E}
}

@Article{Martin-Torres_2005,
       author = {{Martin-Torres}, F.~J. and {Mlynczak}, M.~G.},
        title = "{FUTBOLIN (Full Transfer by Ordinary LINe-by-line methods): a new Radiative Transfer Code for Atmospheric Calculations in the Visible and Infrared}",
    booktitle = {AGU Spring Meeting Abstracts},
         year = 2005,
       volume = {2005},
        month = may,
          eid = {A21A-05},
        pages = {A21A-05},
       adsurl = {https://ui.adsabs.harvard.edu/abs/2005AGUSM.A21A..05M}
}

@article{Gordley_1994,
title = {Linepak: Algorithms for modeling spectral transmittance and radiance},
journal = {Journal of Quantitative Spectroscopy and Radiative Transfer},
volume = {52},
number = {5},
pages = {563-580},
year = {1994},
issn = {0022-4073},
doi = {https://doi.org/10.1016/0022-4073(94)90025-6},
url = {https://www.sciencedirect.com/science/article/pii/0022407394900256},
author = {Larry L. Gordley and Benjamin T. Marshall and D. {Allen Chu}},
}

@Article{Mayer_2005,
AUTHOR = {Mayer, B. and Kylling, A.},
TITLE = {Technical note: The libRadtran software package for radiative transfer calculations - description and examples of use},
JOURNAL = {Atmospheric Chemistry and Physics},
VOLUME = {5},
YEAR = {2005},
NUMBER = {7},
PAGES = {1855--1877},
URL = {https://acp.copernicus.org/articles/5/1855/2005/},
DOI = {10.5194/acp-5-1855-2005}
}

@BOOK{Villanueva_2022,
       author = {{Villanueva}, Geronimo Luis and {Liuzzi}, Giuliano and {Faggi}, Sara and {Protopapa}, Silvia and {Kofman}, Vincent and {Fauchez}, Thomas and {Stone}, Shane Wesley and {Mandell}, Avi Max},
        title = "{Fundamentals of the Planetary Spectrum Generator}",
         year = 2022,
       adsurl = {https://ui.adsabs.harvard.edu/abs/2022fpsg.book.....V}
}

@article{Dudhia_2017,
title = {The Reference Forward Model (RFM)},
journal = {Journal of Quantitative Spectroscopy and Radiative Transfer},
volume = {186},
pages = {243-253},
year = {2017},
note = {Satellite Remote Sensing and Spectroscopy: Joint ACE-Odin Meeting, October 2015},
issn = {0022-4073},
doi = {https://doi.org/10.1016/j.jqsrt.2016.06.018},
url = {https://www.sciencedirect.com/science/article/pii/S0022407316301029},
author = {Anu Dudhia},
}

@misc{Govaerts_2006,
Title = {RTMOM V0B.10 User’s Manual},
year = {2006},
publisher = {EUMETSAT},
author = {Govaerts, Y. M.},
}

@article{Saunders_1999,
author = {Saunders, R. and Matricardi, M. and Brunel, P.},
title = {An improved fast radiative transfer model for assimilation of satellite radiance observations},
journal = {Quarterly Journal of the Royal Meteorological Society},
volume = {125},
number = {556},
pages = {1407-1425},
doi = {https://doi.org/10.1002/qj.1999.49712555615},
year = {1999}
}

@article{Bourassa_2008,
title = {SASKTRAN: A spherical geometry radiative transfer code for efficient estimation of limb scattered sunlight},
journal = {Journal of Quantitative Spectroscopy and Radiative Transfer},
volume = {109},
number = {1},
pages = {52-73},
year = {2008},
issn = {0022-4073},
doi = {https://doi.org/10.1016/j.jqsrt.2007.07.007},
author = {A.E. Bourassa and D.A. Degenstein and E.J. Llewellyn},
}

@Article{Zawada_2015,
AUTHOR = {Zawada, D. J. and Dueck, S. R. and Rieger, L. A. and Bourassa, A. E. and Lloyd, N. D. and Degenstein, D. A.},
TITLE = {High-resolution and Monte Carlo additions to the SASKTRAN radiative transfer model},
JOURNAL = {Atmospheric Measurement Techniques},
VOLUME = {8},
YEAR = {2015},
NUMBER = {6},
PAGES = {2609--2623},
URL = {https://amt.copernicus.org/articles/8/2609/2015/},
DOI = {10.5194/amt-8-2609-2015}
}

@article{Berk_1998,
title = {MODTRAN Cloud and Multiple Scattering Upgrades with Application to AVIRIS},
journal = {Remote Sensing of Environment},
volume = {65},
number = {3},
pages = {367-375},
year = {1998},
issn = {0034-4257},
doi = {https://doi.org/10.1016/S0034-4257(98)00045-5},
url = {https://www.sciencedirect.com/science/article/pii/S0034425798000455},
author = {A. Berk and L.S. Bernstein and G.P. Anderson and P.K. Acharya and D.C. Robertson and J.H. Chetwynd and S.M. Adler-Golden}
}

@article{Fiorino_2014,
      author = "Steven T. Fiorino and Robb M. Randall and Michelle F. Via and Jarred L. Burley",
      title = "Validation of a UV-to-RF High-Spectral-Resolution Atmospheric Boundary Layer Characterization Tool",
      journal = "Journal of Applied Meteorology and Climatology",
      year = "2014",
      publisher = "American Meteorological Society",
      address = "Boston MA, USA",
      volume = "53",
      number = "1",
      doi = "10.1175/JAMC-D-13-036.1",
      pages=      "136 - 156",
      url = "https://journals.ametsoc.org/view/journals/apme/53/1/jamc-d-13-036.1.xml"
}

@article{Clough_2005,
title = {Atmospheric radiative transfer modeling: a summary of the AER codes},
journal = {Journal of Quantitative Spectroscopy and Radiative Transfer},
volume = {91},
number = {2},
pages = {233-244},
year = {2005},
issn = {0022-4073},
doi = {https://doi.org/10.1016/j.jqsrt.2004.05.058},
url = {https://www.sciencedirect.com/science/article/pii/S0022407304002158},
author = {S.A. Clough and M.W. Shephard and E.J. Mlawer and J.S. Delamere and M.J. Iacono and K. Cady-Pereira and S. Boukabara and P.D. Brown},
}

@Article{DeSouza-Machado_2020,
AUTHOR = {DeSouza-Machado, S. and Strow, L. L. and Motteler, H. and Hannon, S.},
TITLE = {kCARTA: a fast pseudo line-by-line radiative transfer algorithm with analytic Jacobians, fluxes, nonlocal thermodynamic equilibrium, and scattering for the infrared},
JOURNAL = {Atmospheric Measurement Techniques},
VOLUME = {13},
YEAR = {2020},
NUMBER = {1},
PAGES = {323--339},
URL = {https://amt.copernicus.org/articles/13/323/2020/},
DOI = {10.5194/amt-13-323-2020}
}

@MISC{Edwards_1992,
       author = {{Edwards}, D.~P.},
        title = "{GENLN2: A general line-by-line atmospheric transmittance and radiance model. Version 3.0: Description and users guide}",
         year = 1992,
        month = jan,
       adsurl = {https://ui.adsabs.harvard.edu/abs/1992ggll.rept.....E}
}

@MISC{Smith_1978,
       author = {{Smith}, H.~J.~P. and {Dube}, D.~J. and {Gardner}, M.~E. and {Clough}, S.~A. and {Kneizys}, F.~X.},
        title = "{FASCODE: Fast Atmospheric Signature CODE (spectral transmittance and radiance)}",
         year = 1978,
        month = jan,
       adsurl = {https://ui.adsabs.harvard.edu/abs/1978fasc.rept.....S}
}

@article{Brogi_2019,
doi = {10.3847/1538-3881/aaffd3},
url = {https://dx.doi.org/10.3847/1538-3881/aaffd3},
year = {2019},
month = {feb},
publisher = {The American Astronomical Society},
volume = {157},
number = {3},
pages = {114},
author = {Matteo Brogi and Michael R. Line},
title = {Retrieving Temperatures and Abundances of Exoplanet Atmospheres with High-resolution Cross-correlation Spectroscopy},
journal = {The Astronomical Journal},
}

@article{Seidel_2020,
	author = {{Seidel}, J. V. and {Ehrenreich, D.} and {Pino, L.} and {Bourrier, V.} and {Lavie, B.} and {Allart, R.} and {Wyttenbach, A.} and {Lovis, C.}},
	title = {Wind of change: retrieving exoplanet atmospheric winds from high-resolution spectroscopy},
	DOI= "10.1051/0004-6361/201936892",
	url= "https://doi.org/10.1051/0004-6361/201936892",
	journal = {A\&A},
	year = 2020,
	volume = 633,
	pages = "A86",
}

@article{Speagle_2020,
    author = {Speagle, Joshua S},
    title = "{dynesty: a dynamic nested sampling package for estimating Bayesian posteriors and evidences}",
    journal = {Monthly Notices of the Royal Astronomical Society},
    volume = {493},
    number = {3},
    pages = {3132-3158},
    year = {2020},
    month = {02},
    issn = {0035-8711},
    doi = {10.1093/mnras/staa278},
    url = {https://doi.org/10.1093/mnras/staa278},
    eprint = {https://academic.oup.com/mnras/article-pdf/493/3/3132/32890730/staa278.pdf},
}

@article{Dave_1964,
  title={Meaning of Successive Iteration of the Auxiliary Equation in the Theory of Radiative Transfer.},
  author={Jitendra V. Dave},
  journal={The Astrophysical Journal},
  year={1964},
  volume={140},
  doi = {10.1086/148024},
  pages={1292}
}

\begin{landscape}
\begin{tiny}
\begin{table}
\setlength{\tabcolsep}{1pt}
\caption{Some radiative transfer and inversion codes used in the (exo) planetary community. Symbols: OE algorithm ($\dagger$), nested sampling ($\ddagger$), MCMC method ($\blacktriangle$), Grid search ($\bullet$), and ML approaches ($\blacklozenge$).}              
\label{table:1}      
\centering 
\begin{tabular}{c p{10cm}  c  }          
\toprule \hline                        
Code name & Link & Reference \\    
\midrule                                 
    Alfnoor $^{\ddagger}$ & - & \cite{Changeat_2020}  \\
    
   APOLLO $^{\blacktriangle}$ & \href{https://github.com/alexrhowe/APOLLO}{\url{https://github.com/alexrhowe/APOLLO}} & \cite{Howe_2017, Howe_2022} \\
   
    ARCiS $^{\ddagger}$ & \href{http://www.exoclouds.com}{\url{http://www.exoclouds.com}} & \cite{Min_2020}  \\
    
    ARTS $^{\dagger}$ & \href{https://radiativetransfer.org/}{\url{https://radiativetransfer.org/}} & \cite{Buehler_2018}       \\

    ASIMUT $^{\dagger}$ & - & \cite{Vandaele_2006a, Vandaele_2006b} \\
    
    ATMO $^\blacktriangle \ddagger \bullet$ & \href{https://www.erc-atmo.eu/?page_id=322}{\url{https://www.erc-atmo.eu/?page\_id=322}} & \begin{tabular}{@{}c@{}}\cite{Tremblin_2015}; \\ \cite{Wakeford_2017}; \\ \cite{Evans_2017}\end{tabular}       \\
    
    AURA $^{\ddagger}$ & - & e.g. \cite{Pinhas_2018} \\
    
    Aura-3D $^{\ddagger}$ & - & \cite{Nixon_2022}     \\

    Aurora $^{\ddagger}$ & - & \cite{Welbanks_2021} \\
    
    BART $^{\blacktriangle}$ & \href{https://github.com/exosports/BART}{\url{https://github.com/exosports/BART}} & \cite{Blecic_2016}       \\

    Benneke \& Seager $^{\ddagger}$ & - & \cite{Benneke_2013}     \\

    Brewster $^{\ddagger \blacktriangle}$ & - & \cite{Burningham_2017}    \\

    Carri\'on-Gonz\'alez et al. $^{\blacktriangle}$ & - & \cite{Carrion-Gonzalez_2020}    \\

    Cerberus $^{\blacktriangle}$ & - & \cite{Swain_2014}  \\
    
    CCF-sequence $^{\blacklozenge}$ & - & \cite{Fisher_2020} \\
    
    CHIMERA $^{\dagger \text{ } \blacktriangle \text{ } \ddagger}$ & \href{https://github.com/mrline/CHIMERA}{\url{https://github.com/mrline/CHIMERA}} & \cite{Line_2013}      \\

    DISORT & 
    \href{http://www.rtatmocn.com/}{\url{http://www.rtatmocn.com/}} & \cite{stamnes2000} \\
    
    Dynesty $^{\ddagger}$ & \href{https://dynesty.readthedocs.io/en/stable/}{\url{https://dynesty.readthedocs.io/en/stable/}} & \cite{Speagle_2020} \\

    Ehrenreich et al. & - & \cite{Ehrenreich_2006} \\

    exoCNN $^{\blacklozenge}$ & \href{https://gitlab.astro.rug.nl/ardevol/exocnn}{\url{https://gitlab.astro.rug.nl/ardevol/exocnn}} & \cite{Ard_vol_Mart_nez_2022}   \\
    
    ExoGAN $^{\blacklozenge}$ & \href{https://osf.io/6dxps/}{\url{https://osf.io/6dxps/}} & \cite{Zingales_2018}  \\

    ExoJAX $^{\blacktriangle}$ & \href{https://github.com/HajimeKawahara/exojax}{\url{https://github.com/HajimeKawahara/exojax}} & \cite{Kawahara_2022} \\

    ExoReL$^{\mathcal{R} \ddagger}$ & - & \cite{Damiano_2020} \\
    
    exoretrievals $^{\ddagger}$ & - & \cite{espinoza_2018} \\

    Exo-REM & \href{https://gitlab.obspm.fr/Exoplanet-Atmospheres-LESIA/exorem}{\url{https://gitlab.obspm.fr/Exoplanet-Atmospheres-LESIA/exorem}} &  \begin{tabular}{@{}c@{}}\cite{Baudino_2015}; \\ \cite{Baudino_2017} \end{tabular} \\

    Fortney et al. $^{\bullet}$ & - & \cite{Fortney_2005, Fortney_2010} \\
    
    gCMCRT & \href{https://github.com/ELeeAstro/gCMCRT}{\url{https://github.com/ELeeAstro/gCMCRT}} & \cite{Lee_2022} \\

\end{tabular}
\end{table}
\newpage
\begin{table}[]
    \centering
    \begin{tabular}{c p{12cm} c }
                     
Code name & Link & Reference \\    
\toprule                               

    Gibson et al.$^{\blacktriangle}$  & - & \cite{Gibson_2020}    \\

    HARP & \href{https://github.com/luminoctum/athena-harp}{\url{https://github.com/luminoctum/athena-harp}} & \cite{Li_2018} \\  
    
    HELA $^{\blacklozenge}$ & \href{https://github.com/exoclime/HELA}{\url{https://github.com/exoclime/HELA}} & \cite{Marquez-Neila_2018} \\

    HELIOS-R $^{\ddagger}$ & \href{https://github.com/exoclime/HELIOS}{\url{https://github.com/exoclime/HELIOS}} & \begin{tabular}{@{}c@{}} \cite{Lavie_2017}; \\ \cite{Oreshenko_2017} \end{tabular} \\
    
    Helios-r2 $^{\ddagger}$ & \href{https://github.com/exoclime/Helios-r2}{\url{https://github.com/exoclime/Helios-r2}} & \begin{tabular}{@{}c@{}}\cite{Kitzmann_2020} \end{tabular}     \\ 

    Home made (MPS) $^{\dagger}$ & - & \cite{Jarchow_1998}       \\

    HRCCS $^{\ddagger}$ & \href{https://www.dropbox.com/sh/0cxfolfmrs8ip37/AABZYoHr8nuRlHJG84dArX4ea?dl=0}{\url{https://www.dropbox.com/sh/0cxfolfmrs8ip37/AABZYoHr8nuRlHJG84dArX4ea?dl=0}} & \cite{Brogi_2019} \\
    
    HyDRA $^{\ddagger}$ & - & \cite{Gandhi_2017}      \\

    HyDRo $^{\ddagger}$ & - & \cite{Piette_2021} \\
    
    INARA $^{\blacklozenge}$ & \href{https://gitlab.com/frontierdevelopmentlab/astrobiology/inara}{\url{https://gitlab.com/frontierdevelopmentlab/astrobiology/inara}} & \cite{Soboczenski_2018}       \\

    Johnson \& Marley $^{\blacklozenge}$ & \href{https://github.com/WreckItTim/MLP-Estimating-Exoplanet-Parameters}{\url{https://github.com/WreckItTim/MLP-Estimating-Exoplanet-Parameters}} & \cite{Johnsen_2020}  \\

    KOPRA & \href{https://www.imk-asf.kit.edu/english/312.php}{\url{https://www.imk-asf.kit.edu/english/312.php}} & \begin{tabular}{@{}c@{}}
    \cite{Stiller_1998a}; \\ \cite{Stiller_2000}; \\ \cite{Stiller_1998b}: \\ \cite{Stiller_2002}\end{tabular} \\

     Lellouch et al. $^{\dagger}$ & - & \cite{Lellouch_2017} \\

     Lupu et al. $^{\blacktriangle \ddagger}$ & - & \cite{Lupu_2016}   \\

    Madhusudhan \& Seager $^{\bullet \blacktriangle}$ & - & \cite{Madhusudhan_2009}     \\

    Madhusudhan et al. $^{\blacktriangle}$ & - &\begin{tabular}{@{}c@{}}\cite{Madhusudhan2010} \\ 
    \cite{Madhusudhan_2011}\end{tabular}  \\

    MARGE $^{\blacklozenge}$ & \href{ https://github.com/exosports/marge}{\url{github.com/exosports/marge}} & \cite{Himes_2022} \\

    Marley \& McKay & - & \cite{MARLEY_1999} \\
    
    MassSpec $^{\blacktriangle}$ & - & \cite{deWit_2013} \\

    MERC $^{\ddagger}$ & - & \cite{Seidel_2020} \\

    METIS $^{\blacktriangle}$ & - & \cite{Lacy_2020} \\
    
    MOLIERE $^{\dagger}$ & - & \cite{urban_2004}       \\

    Moreno et al. & - & \cite{Moreno_1998, Moreno_2001} \\
    
    NEMESIS $^{\dagger \ddagger}$ & \href{https://users.ox.ac.uk/~atmp0035/nemesis.html}{\url{https://users.ox.ac.uk/~atmp0035/nemesis.html}} & \cite{Irwin_2008}     \\

    Nixon \& Madhusudhan $^{\blacklozenge}$ & - & \cite{Nixon_2020}   \\

    PETRA $^{\bullet\text{ } \blacktriangle}$ & - & \cite{Lothringer_2020} \\

\end{tabular}
\end{table}
\newpage
\begin{table}[]
    \centering
    \begin{tabular}{c p{12cm} c }
                     
Code name & Link & Reference \\    
\toprule                               

    petitRADTRANS $^{\ddagger \blacktriangle}$ & \href{https://petitradtrans.readthedocs.io/en/latest/}{\url{https://petitradtrans.readthedocs.io/en/latest/}} & \cite{Mollière_2019}     \\

    PICASO  & \href{https://github.com/natashabatalha/picaso}{\url{https://github.com/natashabatalha/picaso}} & \begin{tabular}{@{}c@{}} \cite{Robbins-Blanch_2022} \\ \cite{Batalha_2019} \end{tabular} \\
    
    plan-net $^{\blacklozenge}$ & \href{ https://github.com/exoml/plan-net}{\url{ https://github.com/exoml/plan-net}} & \cite{Cobb_2019}  \\
    
    PLATON $^{\ddagger}$ & \href{https://github.com/ideasrule/platon}{\url{https://github.com/ideasrule/platon}} & \cite{Zhang_2020}       \\

    PLATON II $^{\ddagger}$ & \href{https://github.com/ideasrule/platon}{\url{https://github.com/ideasrule/platon}} & \cite{Zhang_2020b} \\

    PolyChord $^{\ddagger}$ & \href{https://github.com/PolyChord/PolyChordLite}{\url{https://github.com/PolyChord/PolyChordLite}} & \cite{Handley_2015a, Handley_2015b} \\
    
    POSEIDON $^{\ddagger}$ & \href{https://github.com/MartianColonist/POSEIDON}{\url{https://github.com/MartianColonist/POSEIDON}} & \cite{MacDonald_2017}    \\
    
    PSG $^{\dagger \ddagger}$ & \href{https://github.com/nasapsg}{\url{https://github.com/nasapsg}} & \cite{Villanueva_2018}       \\

    Pyrat-Bay $^{\blacktriangle}$ & \href{https://pyratbay.readthedocs.io/en/latest/}{\url{https://pyratbay.readthedocs.io/en/latest/}} & \cite{Cubillos_2021} \\

    Pytmosph3R  & \href{https://pypi.org/project/pytmosph3r/}{\url{https://pypi.org/project/pytmosph3r/}} & \cite{Caldas_2019, Falco_2022} \\

    p-winds$^{\blacktriangle}$ & \href{https://github.com/ladsantos/p-winds}{\url{https://github.com/ladsantos/p-winds}} & \cite{Dos_Santos_2022}   \\

    REDFOX & - & \cite{Scheucher2020} \\

    rfast$^{\blacktriangle}$ & \href{https://github.com/hablabx/rfast}{\url{https://github.com/hablabx/rfast}} & \cite{Robinson_2023}    \\

    SCARLET $^{\ddagger \blacktriangle}$ & - & \cite{Benneke_2015}       \\

    smarter $^{\ddagger}$ & - & \cite{Lustig-Yaeger_2022} \\

    species $^{\ddagger \bullet \blacktriangle}$ & \href{https://github.com/tomasstolker/species}{\url{https://github.com/tomasstolker/species}} & \cite{Stolker_2020}   \\
    
    Tau & \href{https://cpc.cs.qub.ac.uk/summaries/AEPN_v1_0.html}{\url{https://cpc.cs.qub.ac.uk/summaries/AEPN_v1_0.html}} & \cite{Hollis_2013} \\
    
    TauREx 2.6 $^{\ddagger \blacktriangle}$ & \href{https://github.com/ucl-exoplanets/TauREx_public}{\url{https://github.com/ucl-exoplanets/TauREx\_public}} & \begin{tabular}{@{}c@{}} \cite{Waldmann_2015a}; \\ \cite{Waldmann_2015b} \\
    \end{tabular} \\
    
    TauREx 3.1 $^{\ddagger}$& \href{https://taurex3-public.readthedocs.io/en/latest/}{\url{https://taurex3-public.readthedocs.io/en/latest/}} & \cite{al-refaie_2021}       \\

    ThERESA $^{\blacktriangle}$ & - & \cite{Challener_2022}   \\
    
    tierra $^{\blacktriangle}$ & \href{https://github.com/disruptiveplanets/tierra}{\url{https://github.com/disruptiveplanets/tierra}} & \cite{Niraula_2022} \\

    TRIDENT  & - & \cite{MacDonald_2022} \\

    Vasist et al. $^{\blacklozenge}$ & - & \cite{vasist_2023} \\

    VI-retrieval $^{\blacklozenge}$ & - & \cite{yip_2022} \\

\bottomrule                                            
\end{tabular}
\end{table}
\end{tiny}
\end{landscape}

\end{document}